On the solution and Lagrangian representation of Duffing oscillator with damping

Amitava Chaudhuri [1], Madan Mohan Panja [2] Benoy Talukdar [3]

[1] Department of Physics, The University of Burdwan University, Purba Bardhamann 713104 , West Bengal, India

[2] Department of Mathematics, Visva-BharatI University, Santiniketan 731235, West Bengal, India

[3] Department of Physics, Visva-Bharati University, Santiniketan 731235, West Bengal, India

**Abstract**. We construct the equation of Duffing oscillator in a dissipative medium using certain concepts from elementary mechanics. The Duffing equation (DE) without damping can be solved analytically. This is not true for a DE that involves a damping term. We remove the damping term from a linearly damped DE and thus obtain a simple analytical solution $x(t)$ of the damped Duffing equation in the weak damping limit. The constructed solution allows us to examine the effect of damping on the phase path of the oscillator. The phase path is a parametric plot of $x(t)$ and $\dot{x}(t)(=dx(t)/dt)$ on the plane $(x(t),\dot{x}(t))$. While the phase path of the un-damped Duffing oscillator is an isolated limit cycle, the corresponding phase path for the Duffing oscillator with damping is a distorted one. We confirm our observation on the effect of dissipation by numerical simulation. We point out that it is often of interest to study dissipative systems at the quantum level and construct Lagrangian representations for both un-damped and damped Duffing oscillators. These results are expected to play a role to quantize the systems. We make some additional comments in respect of this.

1.  **Introduction**

It is widely believed that linear elastic springs which obey Hooke's law are a mainstay of elementary mechanics [1]. If a spring of length $L$ fixed at one end is pulled by an external force $F$ to increase $L$ by an amount $x$, Hooke's law states that the force is proportional to the extension such that

$$F = kx. \qquad (1)$$

The constant of proportionality, $k$, is called the spring constant and measures the stiffness of the spring. In eq.(1) both $F$ and $x$ are vectors having the same direction. In the present work we deal with a one dimensional system and as such avoid use of vector sign. Newton's third law of motion implies that the spring must act back with a force $-kx$. If a mass $m$ attached to the free end of the spring is displaced by a small amount along the $x$ axis and then released, it will vibrate or oscillate back and forth about an equilibrium position. In that case $x$ will become a function of time $t$. However, for clarity of presentation we shall use $x$ instead of $x(t)$. The argument in $x$ will be used only if it is absolutely necessary for clarity of presentation. Newton's second law in conjunction with Hooke's law gives the equation of motion

$$m\ddot{x} = -kx, \quad \ddot{x} = \frac{d^2x}{dt^2} \tag{2}$$

for the vibrating system. As in eq.(2) we shall henceforth use over dots to denote differentiation with respect to $t$. If motion of the mass takes place in a viscous/dissipative medium we should add a damping force $b\dot{x}$ to left side of eq.(2) and thus write a modified equation of motion

$$m\ddot{x} + b\dot{x} + kx = 0, \quad \dot{x} = \frac{dx}{dt}, \quad b > 0. \tag{3}$$

If we make the spring nonlinear, eq.(3) will further modify to

$$m\ddot{x} + b\dot{x} + kx + \alpha x^3 = 0, \quad \alpha > 0. \tag{4}$$

In writing eq.(4) we have considered only the cubic nonlinearity. We could also consider the quintic or even a combination of cubic and quintic nonlinearities. For $m = \alpha = 1$, $k = -1$ and $b = 2\gamma$, eq.(4) reduces to the well known equation of the Duffing-oscillator with damping

$$\ddot{x} + 2\gamma\dot{x} - x + x^3 = 0, \tag{5}$$

where $\gamma$ stands for the coefficient of friction of the viscous medium in which the system is embedded. One may reasonably ask : why in eq.(5) we have used $b = 2\gamma$ rather than $b = \gamma$. We shall see in the course of our study that the present choice help us display the solution of eq.(5) in a rather neat form.

Mechanistically, the Duffing equation **without the damping term** describes the motion of a classical particle in a double-well potential

$$V(x) = -\frac{x^2}{2} + \frac{x^4}{4} \tag{6}$$

as shown in figure 1. In presenting the curve in fig.1 we have chosen units of length so that the minima are at $x = \pm 1$, and the units of energy so that the depth of each well is at $-1/4$. Interestingly, the Duffing equation involving dissipative term $2\gamma\dot{x}$ can be used as a convenient mathematical model for a wide variety of physical systems [2].

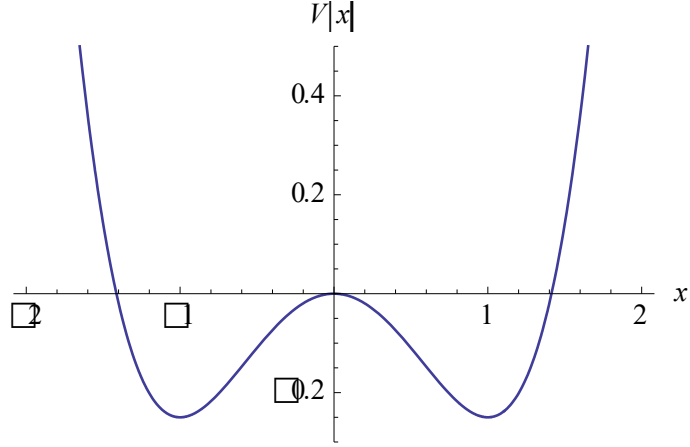

Figure 1. Potential function $V(x)$ as a function of $x$.

In the travelling wave coordinate many partial differential equations appearing in soliton theory [3] transform to Duffing-type equations. This fact has recently been discussed in some detail by Salas et. al [4]. It is remarkable to note that the nonlinear Schrodinger equation

$$i\frac{\partial u}{\partial t} + \frac{\partial^2 u}{\partial x^2} + \lambda |u|^2 u = 0, \quad u = u(x,t) \text{ and } \lambda = \text{constant} \tag{7}$$

which plays a crucial role in nonlinear optics can be reduced in the form of Duffing equation

$$\frac{d^2 v(\xi)}{d\xi^2} - (\alpha^2 + \beta^2) v(\xi) + 2\lambda v^3(\xi) = 0 \tag{8}$$

using the transformation

$$u(x,t) = e^{i(\alpha x + \beta t)} v(\xi) \tag{9}$$

with the travelling coordinate $\xi = x - \alpha t$.

The un-damped Duffing equation obtained from eq.(5) using $\gamma = 0$ can be solved analytically. On the other hand, the corresponding equation involving the effect of dissipation can be treated only numerical methods. Consequently, there exists a vast amount of literature for constructing approximate analytic solution for the Duffing equation with damping. See, for example, the work by Nourazar and Mirzabeigy [5] and references therein. While the authors in ref.5 made use of a modified differential transform method to provide an approximate analytic solution of the problem, Johannessen [6] found an exact analytic solution of the damped Duffing equation under the constraint that the damping term and initial conditions of the system satisfy an algebraic equation. This implies that the solution presented is specific to restricted type of damping. In the present work we are interested to provide a general solution of the linearly damped Duffing-oscillator equation for small values of $\gamma$ and make use of it to investigate the effect of dissipation on the dynamics of the system. We shall achieve this by removing the first derivative or damping term from the equation by using a transformation that allows studying the dissipative system in a non-dissipative framework [7]. There is a standard procedure [8] to remove the first derivative term from a linear ordinary differential equation and thus convert a non-self adjoint equation to the self-adjoint form. This is, however, not so obvious for nonlinear differential equations [9]. In addition to providing analytical solutions for Duffing equations (both un-damped and damped), we shall also present results for their Lagrangians [10]. The Lagrangian formulation of physical problems provides methods by which theories are developed for quantum mechanics and modern physics [11]. In the following we briefly outline the method we follow to find Lagrangianss.

In the calculus of variation there are two types of problems. The first one, often called the direct problem, is fairly straightforward. Here one uses the Lagrangian function in the Euler-Lagrange equation [10] to write the equation of motion. On the other hand, the second one is more complicated and consists in constructing the Lagrangian function from the equation of motion. This is the inverse problem of the calculus of variation [12]. The set of necessary and sufficient conditions for the existence of Lagrangians for Newtonian equations is provided by the so-called Helmholtz conditions [13]. These conditions are violated in dissipative systems**.** This might be the reason why Lanscoz [14] believed that forces of friction are outside the realm of vaiational principle such that any equation of motion involving damping term cannot have a Lagrangian representation. However, to look for a Lagrangian representation of Duffing oscillator with damping we shall work with the transformed equation of the system provided by the approach followed in ref.7 in which the dissipative term disappears from the equation of motion. The Lagrangian of this equation is then found by using a method which is particularly suitable for the inverse problem of nonlinear equations [15].

In section 2 we obtain a simple analytic solution of the un-damped Duffing equation and show that the associate Poincare map [16] i.e. the parametric plot of $x(t)$ and $\dot{x}(t)$ on a plane with Cartesian $(x(t), \dot{x}(t))$ is a limit cycle. We also study the time evolution of the system by plotting $x(t)$ and $\dot{x}(t)$ as functions of time, and make some comments regarding the role of nonlinearity in the dynamics of the oscillator. Here we also construct a Lagrangian representation of the system. Section 3 contains the main results of our investigation. The method derived in ref.7 is used to remove the dissipative term from eq.(5) and write a transformed equation in the small damping limit, which is identical to the un-damped equation treated in section 2. Understandably, the equation in the transformed frame can be solved by the same method as used for the un-damped oscillator. The solution of the Duffing system with damping is then constructed from the solution of the transformed equation. We then use our analytic solution to examine in some detail how dissipation affects the phase portrait (Poincare map) and time evolution of $x(t)$ and $\dot{x}(t)$ the damped oscillator. Where-ever possible, we compare the results of our analytical model with corresponding results obtained by numerical calculations. In close analogy with the Lagrangian of the non-dissipative system, an expression for the Lagrangian of the Duffing oscillator with damping is presented. We venture to suggest that the Lagrangian presented will be useful to quantize the damped Duffing sysrem, Finally, in section 4 we summarize our outlook on the present work and make some concluding remarks.

2. **Un-damped Duffing oscillator**

In the absence of damping ($\gamma = 0$) the Duffing oscillator is represented by a simple autonomous differential equation $\ddot{x} - x + x^3 = 0$ which can be integrated to write

$$\frac{dx}{\sqrt{x^2 - \frac{1}{2}x^4 + c_1}} = dt, \qquad (10)$$

where $c_1$ is a constant and is equal to $1/2$ for the initial condition $x(0) = \dot{x}(0) = 1$. On integration, eq.(10) gives

$$-i\sqrt{2a}F(y,k) = t + c_2 \qquad (11)$$

with $c_2$, a new constant of integration to be determined again from the initial conditions. Here $a = \sqrt{2} - 1$. In eq.(11) $F(y,k)$ stands for an elliptic integral of the first kind [17] with

$$y = i \arcsin h(x/a) \text{ and } k = -1 + 2a. \qquad (12)$$

Equation(11) can be solved to get

$$x = -\sqrt{a}isn[(2+\sqrt{2})\sqrt{a}c_2 + 2\sqrt{a}t, k] \ . \qquad (13)$$

Here $sn(.)$, the so called sine-amplitude Jacobi elliptic function [18]. Applying the condition $\dot{x}(0) = 1$ on the derivative of the solution in eq.(13) we get $c_2 = 1$. Thus from eq.(13) we obtain

$$\dot{x} = \frac{1}{\sqrt{2}} cn(\xi(1+t), k) dn(\xi(1+t), k) , \qquad (14)$$

where

$$dn(u,k) = \sqrt{1 - k^2 sn^2 u} , u = \xi(1+t) \text{ with } \xi = i(2+\sqrt{2})\sqrt{a}/2. \qquad (15)$$

The Poincare map represents a geometrical device to extract, from differential equations, properties of dynamical systems [19] such as periodicity, growth, stability, and so on. In view of this we present in figure 2 the Poincare map generated by the results in eq.(13) and eq.(14). Here we see that as $t$ increases $(x(t), \dot{x}(t))$ traces out a curve on this plane, This curve is called the phase path.

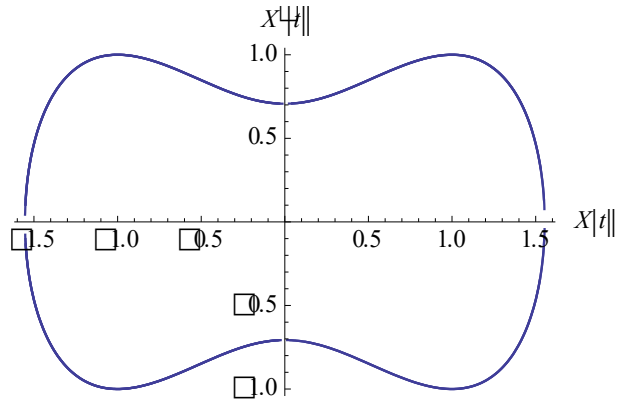

Figure 2. Phase plane of the Duffing oscillator in the absence of damping.

The phase path in figure 2 is an isolated closed curve: isolated in the sense that there is no other closed path in its immediate neighborhood. An isolated closed curve of this type is called a limit cycle. Limit cycles can occur in only nonlinear systems. More specifically, the curve in this figure is a stable limit cycle since if the system is slightly disturbed from its regular oscillatory state, the resulting new path, on either side will be attracted back to the curve,

We portray in figure 3 $x(t)$ and $\dot{x}(t)$ as a function of time. The solution $x(t)$ as a function of $t$ (solid curve) is seen to exhibit oscillatory behavior. In the absence of the cubic term the Duffing equation reduces to $\ddot{x}(t) - x(t) = 0$, the solution of which grows/decays exponentially. Thus it if evident that oscillatory behavior of the solution arises due the presence of the nonlinear term in the equation. . The curve for $\dot{x}(t)$ (dashed curve), although oscillatory in nature, is characterized by small depressions on the crests and troughs of the curve. However, it seems that it has so far remained un-noticed that nonlinearity induces oscillation in a Duffing system.

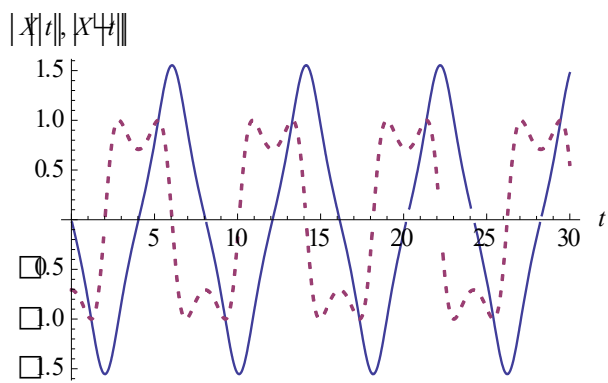

Figure 3. The solution $x(t)$ and its derivative $\dot{x}(t)$ as function of time $t$ (un-damped system)

In the next section we shall study in some detail the effects of dissipation on the dynamics of Duffing oscillator with special attention to phase path and time evolution of the primitive of eq.(5). Meanwhile, we make use of the method of Lopez [15] to provide a simple Lagrangian representation of the Duffing oscillator without damping. Here one proceed by making use of the Jacobi integral [10]

$$\sum_{i=1}^{N} v^i \left( \frac{\partial L}{\partial v^i} \right) - L = K, \tag{16}$$

where $L = L(\overline{x}, \overline{v})$ is the Lagrangian and $K = K(\overline{x}, \overline{v})$ stands for the constant of the motion of a $N$-dimensional second-order ordinary differential equation

$$\ddot{x}^i = f^i(x^j, \dot{x}^j), \quad i, j = 1, 2 \ldots \ldots N. \tag{17}$$

Writing eq.(17) in the equivalent form $\dfrac{dv^i}{dt} == f^i(\overline{x}, \overline{v})$, one can demand that the constant of the motion $K(\overline{x}, \overline{v})$ is a first integral of the equation provided

$$\sum_{i=1}^{N} \left[ f^i(\overline{x}, \dot{v}) \frac{\partial K}{\partial v^i} + v^i \frac{\partial K}{\partial x^i} \right] = 0. \tag{18}$$

The solution of eq.(18) or the integral surface can now be obtained from the equation of characteristic [20]

$$\frac{dv^1}{f^1(\overline{x}, \overline{v})} = \ldots \ldots = \frac{dv^N}{f^N(\overline{x}, \overline{v})} = \frac{dx^1}{v^1} = \ldots \ldots = \frac{dx^N}{v^N} = \frac{dK}{0}. \tag{19}$$

The last term in eq.(19) merely indicates that it can be used to find the general solution of eq.(18), which represents an equation for the constant of the motion of our second-order differential equation (17). For a N-dimensional Newtonian system the solution of eq.(16) can be expressed as an integral over the possible constants of the motion so that [15]

$$L(\overline{x}, \overline{v}) = \frac{1}{N} \sum_{i=1}^{N} v^i \int^{v^i} \frac{K^i(\overline{x}, \xi)}{\xi^2} d\xi. \tag{20}$$

The above Lagrangian provides a solution of the inverse variational problem for the second-order differential equation (17) irrespective of whether it is linear or nonlinear. For the one dimensional system of our interest eq.(20) reads

$$L(x, \dot{x}) = \dot{x} \int^{\dot{x}} \frac{K(x, \xi)}{\xi^2} d\xi. \tag{21}$$

From the equation of the non-dissipative Duffing oscillator it is straightforward to write

$$K(x, \xi) = \frac{\xi^2}{2} - \frac{x^2}{2} + \frac{x^4}{4}. \tag{22}$$

From Eqs(21) and (22) we find the desired Lagrangian as

$$L(x, \dot{x}) = \frac{1}{2}\dot{x}^2 + \frac{1}{2}x^2 - \frac{1}{4}x^4. \tag{23}$$

In view of eq.(6) we see that the above Lagrangian has come out in the well known form $L = T - V$.

3. **Duffing oscillator with Damping**

The non-dissipative Duffing-oscillator equation could be solved analytically. But in the presence of dissipation one needs to use numerical routines to solve the equation of motion for the Duffing system. Here the first-derivative term in the equation is the source of trouble. We make use of a non-point transformation [7,21] to remove the problematic term from the equation. This permits us to obtain an analytic solution of the damped Duffing oscillator similar to that found for the un-damped system. The constructed solution can easily be used to examine effect of dissipation on the Duiffing system. An important aspect of our approach is that the equation in the transformed frame provides a basis to find the Lagrangian representation of the Damped system.

To illustrate our approach to solve eq.(5) consider dissipative equation written in the form

$$\ddot{x} + g(x, \dot{x})\dot{x} + h(x) = 0 \tag{24}$$

The explicit forms of the functions $g(x, \dot{x})$ and $h(x)$ determine nature of the physical system represented by eq.(24). For example, choosing $g(x, \dot{x}) = 2\gamma$ and $h(x) = x$ we get the equation for the damped harmonic oscillator

$$\ddot{x} + 2\gamma\dot{x} + x = 0. \tag{25}$$

On the other hand, if we take $g(x, \dot{x}) = 3\gamma x$ and $h(x) = \gamma^2 x^3 + kx$ we get the so-called generalized Emden-type equation [22]

$$\ddot{x} + 3\gamma x \dot{x} + \gamma^2 x^3 + kx = 0. \tag{26}$$

In fact, we can write a large of equations which involve the first derivative of $x$. However, let us now examine how the transformation [7]

$$y(t) = f(t) e^{\int g(x,\dot{x}) dt} \tag{27}$$

is useful to remove the term involving $\dot{x}$ from any given equation of motion. If we choose $f(t) = x(t)$ and $g(x, \dot{x}) = \gamma$, the transformation in eq.(27) which now reads $y(t) = x(t)e^{\gamma t}$ reduces the damped harmonic oscillator, eq.(25), to a harmonic oscillator equation

$$\ddot{y} + (1 - \gamma^2)y = 0 \tag{28}$$

with $\omega = (1 - \gamma^2)^{\frac{1}{2}}$ as the angular frequency of the vibrating system as viewed in the transformed frame. Similarly, the choice $f(t) = 1$ and $g(x, \dot{x}) = \gamma x(t)$ transform the nonlinear equation (26) to

$$\ddot{y} + ky = a \tag{29}$$

with $a$ an arbitrary constant. There can be many more similar examples.

It is straightforward to write the solution and Lagrangian of eq.(28) as

$$y(t) = \sin(\sqrt{1 - \gamma^2} t) \tag{30}$$

and
$$L = \frac{1}{2}\left(\dot{y}^2 - (1 - \gamma^2)y^2\right) \tag{31}$$

Making use of the mapping $y(t) = x(t)e^{\eta}$ we can write eq.(30) and eq.(31) as

$$x(t) = e^{-\eta} \sin(\sqrt{1 - \gamma^2} t) \tag{32}$$

and
$$L = \frac{1}{2}\left(\dot{x}^2 + 2\gamma x\dot{x} + 2\gamma^2 x^2 - x^2\right)e^{2\eta}. \tag{33}$$

Equation (32) demonstrates the basic feature of wave propagation in a dissipative medium; the wave decays rapidly in media of large damping coefficient. The Lagrangian in eq.(33) when substituted in the Euler-Lagrange equation $\frac{d}{dt}\left(\frac{\partial L}{\partial \dot{x}}\right) - \frac{\partial L}{\partial x} = 0$ gives the equation of motion for the damped harmonic oscillator, namely, eq.(25). A result similar to that in eq.(33) for the Lagrangian of the damped harmonic oscillator was found independently by Caldorila [23] as well as by Kanai [24] quite a long ago. But their result is not connected to that of ours by a Gauge term. Thus our result for $L$ in eq.(33) is of pedagogic importance It is interesting to note that the same transformation, $x(t) = y(t)e^{-\eta}$, as used for the damped harmonic oscillator when applied on the equation of the Duffing oscillator with damping removes the first derivative term from eq.(5) and gives

$$\ddot{y} - (1 + \gamma^2)y + e^{-2\eta} y^3 = 0. \tag{34}$$

We now focus our attention on a Duffing oscillator embedded in a low viscous medium ($\gamma \ll 1$) such that $e^{-2\eta} = 1$. In this case eq.(34) reduces to

$$\ddot{y} - (1 + \gamma^2)y + y^3 = 0 \tag{35}$$

Since the equation in (35) does not involve $\dot{y}$, it can be solved by following the same approach as was used to find the primitive of the un-damped Duffing oscillator. From the solution of eq.(35) we obtain the solution of the Duffing oscillator with Damping as

$$x(t) = -ib_-^{1/2} e^{-\eta} sn(it(b_+/2)^{1/2}, -b_-/b_+) \qquad (36)$$

with

$$b_\pm = (2 + 2\gamma^2 + \gamma^4)^{1/2} \times (1+\gamma^2). \qquad (37)$$

We have verified that for $\gamma = 0$ the result in eq.(36) goes over to that for the un-damped oscillator given in eq.13 and reproduces the phase diagram in figure 2 as well as the curves for $x(t)$ and $\dot{x}(t)$ in figure 3. Let us now make use of the solution in eq.(36) and its time derivative to examine the effects of dissipation on the phase diagram and associated time evolution of $x(t)$ and $\dot{x}(t)$ for different values of $\gamma$. The observed changes in these diagrams will be compared with corresponding portraits obtained by numerical calculation.

Figure 4. shows the phase plane for $\gamma = 0.01$ obtained using our analytic expression in eq.(36). The corresponding phase plane found from numerical solution of eq.(5) is presented in figure 5. To solve eq.5 numerically we wrote it as a system of equations, namely,

$$\dot{x}(t) = v(t) \qquad (38a)$$

and

$$\dot{v}(t) = -2\gamma\dot{x}(t) + x(t) - x(t)^3, \qquad (38b)$$

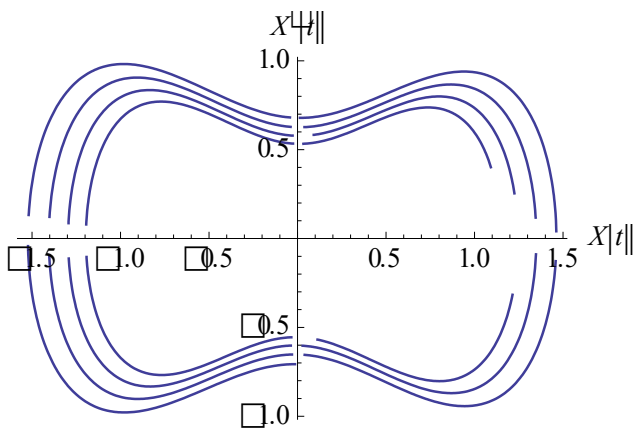

Figure 4. Phase plane of the damped Duffing oscillator for $\gamma = 0.01$ (Analytic model).

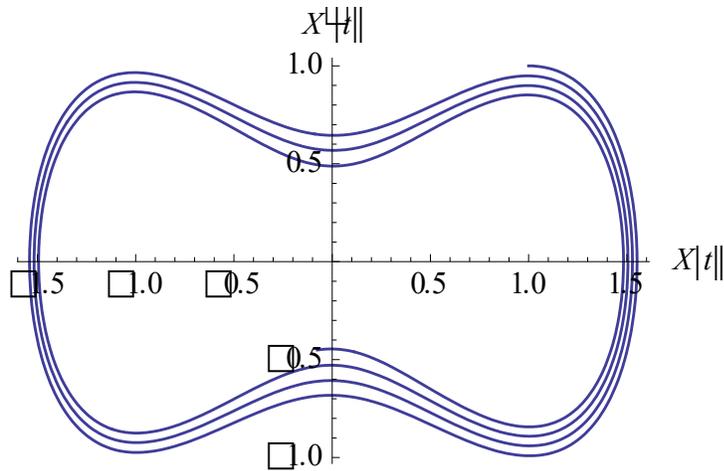

Figure 5. Phase plane of the damped Duffing oscillator for $\gamma = 0.01$ (Numerical calculation of eq.(5)).

and solved them using Mathematica with the initial condition $x(0) = \dot{x}(0) = 1$. Note that we used the same initial condition to write the solutions in eq.(13) and eq.(36). We have seen in figure 2 that the phase plane of the Duffing oscillator in the absence of damping is an isolated closed path. Comparing the curves in figures 2 and 4 we see that the mean limit cycle of the un-damped oscillator is now distorted due to dissipative effects. This fact is supported by the curve in figure 5 drawn by using a purely numerical routine. In the recent past, similar distortion of the mean limit cycle was studied in a stochastically driven nonlinear oscillator [25]. We display in figure 6 $x(t)$ and $\dot{x}(t)$ for the same $\gamma$ value as a function of time. The curves in this figure closely resemble the corresponding curve in fig.3.

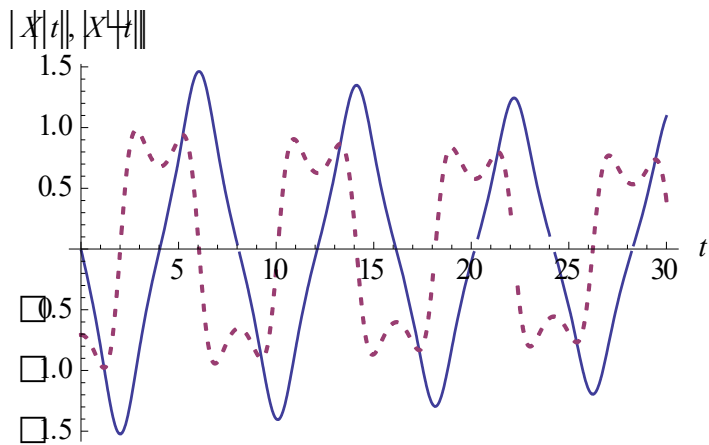

Figure 6. Plot of $x$ and $\dot{x}$ as a function of time (Analytical model)

In both figures the undulations in the curves for $x(t)$ and $\dot{x}(t)$ look alike. However, closely looking into these figures we see that, as opposed to the curves in figure 3, the amplitudes in the corresponding curves for the damped system (figure 6) decrease with time. Understandably, the diminution in amplitudes, whatsoever, has its physical origin to the dissipation of energy from the oscillator to the surrounding medium. Figure 7 displays the curves for $x(t)$ and $\dot{x}(t)$ obtained using numerical routine.

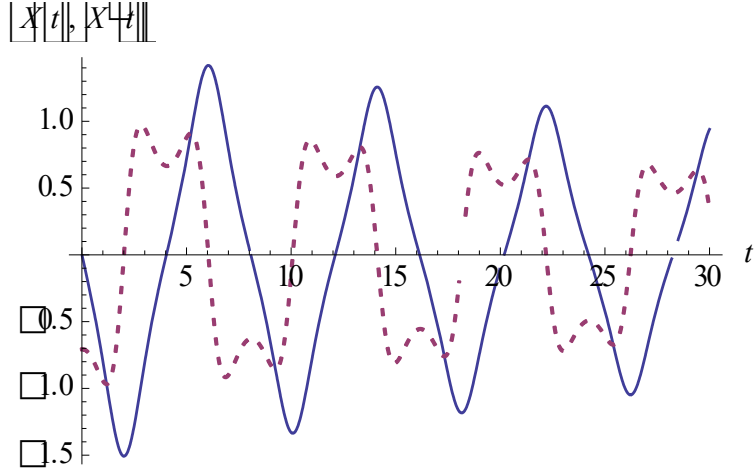

Figure 7. Plot of $x$ and $\dot{x}$ as a function of time (Numerical calculation of eq.(5))

Curves in this figure are similar to the corresponding curves in figure 6. It is expected that for higher values of $\gamma$ the phase plane will be more distorted and simultaneously $x(t)$ and $\dot{x}(t)$ will decay more rapidly with time. We have verified that this is indeed the case. We now find the Lagrangian representation for the Duffing equation with damping.

From eq.(35) it is straightforward to write

$$K(x,\xi) = \frac{\xi^2}{2} - \frac{1}{2}(1+\gamma^2)y(t)^2 + \frac{y(t)^4}{4}. \tag{39}$$

Substitution of the expression in eq.(39) in eq.(21) gives

$$L = \frac{1}{2}\dot{y}(t)^2 + \frac{1}{2}(1+\gamma^2)y(t)^2 - \frac{1}{4}y(t)^4 \tag{40}$$

for the Lagrangian of Damped Duffing equation in the non-dissipative frame. The relation between $y(t)$ and $x(t)$ when used in eq.(40) leads to

$$L = \frac{1}{2}e^{2\gamma t}\left(\dot{x}^2 + 2\gamma x\dot{x} + (1+2\gamma^2)x^2 - \frac{1}{2}e^{2\gamma t}x^4\right), \tag{41}$$

which for $\gamma = 0$ gives the Lagragian for the un-damped equation in eq.(23). Using the result for $L$ from eq.(41) in the Euler-Lagrange equation we find the equation of motion

$$\ddot{x} + 2\gamma\dot{x} - x + e^{2\gamma t}x^3 = 0. \tag{42}$$

for the Duffing Oscillator with damping. If we restrict ourselves to the same physical consideration as applied in writing eq.(35) from eq.(34), the above equation of motion goes over to that in eq.(5).

There exist well established procedures for quantization of conservative systems. But quantization of dissipative systems has always been intriguing because of the intricate nature of the dissipative interaction as well as the continuous time-varying energy of dissipating systems. Thus it will be a highly involved problem to derive an unambiguous quantization procedure for the Duffing oscillator with damping. Currently, we have an efficient mathematical formalism often called the 'Quantum-state diffusion model' [26] to quantize dissipative systems. . This model has been used [27] to discuss the crossover from classical to quantum behavior of the damped Duffing oscillator. But we feel that, in close analogy with the work of Gerry [28] on the quantization of damped harmonic oscillator, the time-dependent Lagrangian in eq.(41) will help us quantize the weakly dissipative Duffing oscillator by the path integral quantization procedure [29].

4. **Concluding remarks**

In this work we studied the dynamics of dissipative Duffing oscillator by going over to a coordinate frame in which the damping term disappears from the equation of motion of the system. This permitted us to obtain an analytic solution for the damped oscillator, which closely resembles the solution of the Duffing equation without damping. Subsequently, we made use of this solution to examine the effect of damping on the properties of the oscillator. The analysis presented by us is based on eq.(35) which is valid only for small values of the damping coefficient $\gamma$. Thus it remains an interesting curiosity to look for a coordinate transformation which not only removes the first derivative term from the Duffing equation with damping but also leads to an equation that can be solved for arbitrary values of $\gamma$.

We solved the inverse variational problem for both non-dissipative and dissipative Duffing equations to find their Lagrangian representations with the hope that these might be of some use to quantize the Duffing oscillator. The Lagrangian of the equation without damping term is time independent such that the corresponding Hamiltonian provides a natural basis to quantize the un-damped Duffing oscillator by the canonical quantization procedure [11.The dissipative system, being characterized by a time-dependent Lagrangian, will require the use of the path integral formulation of quantum mechanics [29]. Recently, an interesting approach to quantization of the damped harmonic oscillator has been derived by Deguchi and Fujiwara [30] on the basis of some modification of the Bateman Lagrangian [31]. The time independent Lagrangian, $L = \dot{x}\dot{y} + \gamma(x\dot{y} - \dot{x}y) - xy$, of Bateman provides an indirect analytic representation of the damped oscillator in the sense that the Euler-Lagrange equation in variables $y$ and $\dot{y}$ gives the dissipative equation written in terms of $x$ and $\dot{x}$. A result for the Bateman-type or indirect analytic representation for the dissipative Duffing oscillator is now available [32]. The constructed time-independent Lagrangian of this work can be profitably used in the approach of ref.30 to quantize Duffing system in dissipative media.